\documentclass{aastex631}%{manuscript} %,twocolumn

\usepackage{amsmath}
\usepackage{amssymb}
\usepackage{graphicx}
\usepackage{CJKutf8}
\begin{document}
\begin{CJK}{UTF8}{gbsn}
%%
%% \documentclass[twocolumn,linenumbers,trackchanges]{aastex631}

%% \hypersetup{linkcolor=red,citecolor=green,filecolor=cyan,urlcolor=magenta}

%% If you want to create your own macros, you can do so
%% using \newcommand. Your macros should appear before
%% the \begin{document} command.
%%
%\newcommand{\vdag}{(v)^\dagger}
%\newcommand\aastex{AAS\TeX}
%\newcommand\latex{La\TeX}

%% Reintroduced the \received and \accepted commands from AASTeX v5.2
%\received{}
%\revised{}
%\accepted{\today}

%% Command to document which AAS Journal the manuscript was submitted to.
%% Adds "Submitted to " the argument.
%\submitjournal{}

%\usepackage{booktabs}
%\usepackage{deluxetable}

\title{Spectral Features of the Solar Transition Region and Chromospheric Lines at Flare Ribbons Observed with IRIS}

\correspondingauthor{M. D. Ding}
\email{dmd@nju.edu.cn}
\correspondingauthor{Y. Li}
\email{yingli@pmo.ac.cn}

\author[0000-0002-0156-5180]{L. F. Wang}
\affiliation{School of Astronomy and Space Science, Nanjing University, Nanjing 210023, People’s Republic of China}
\affiliation{Key Laboratory of Modern Astronomy and Astrophysics (Nanjing University), Ministry of Education, Nanjing 210023, People’s Republic of China} 

\author[0000-0002-8258-4892]{Y. Li}
\affiliation{Key Laboratory of Dark Matter and Space Astronomy, Purple Mountain Observatory, Chinese Academy of Sciences, Nanjing 210023, People’s Republic of China}
\affiliation{School of Astronomy and Space Science, University of Science and Technology of China, Hefei 230026, People's Republic of China}

\author[0000-0001-7540-9335]{Q. Li}
\affiliation{Key Laboratory of Dark Matter and Space Astronomy, Purple Mountain Observatory, Chinese Academy of Sciences, Nanjing 210023, People’s Republic of China}
\affiliation{School of Astronomy and Space Science, University of Science and Technology of China, Hefei 230026, People's Republic of China}

\author[0000-0003-2837-7136]{X. Cheng}
\affiliation{School of Astronomy and Space Science, Nanjing University, Nanjing 210023, People’s Republic of China}
\affiliation{Key Laboratory of Modern Astronomy and Astrophysics (Nanjing University), Ministry of Education, Nanjing 210023, People’s Republic of China} 

\author[0000-0002-4978-4972]{M. D. Ding}
\affiliation{School of Astronomy and Space Science, Nanjing University, Nanjing 210023, People’s Republic of China}
\affiliation{Key Laboratory of Modern Astronomy and Astrophysics (Nanjing University), Ministry of Education, Nanjing 210023, People’s Republic of China}

\begin{abstract}
We report on the spectral features of the Si IV 1402.77 \AA, C II 1334.53 \AA, and Mg II h or k lines, formed in the layers from the transition region to the chromosphere, in three two-ribbon flares (with X-, M-, and C-class) observed with IRIS. All the three lines show significant redshifts within the main flare ribbons, which mainly originate from the chromospheric condensation during the flares. The average redshift velocities of the Si IV line within the main ribbons are 56.6, 25.6, and 10.5 km s$^{-1}$ for the X-, M-, and C-class flares, respectively, which show a decreasing tendency with the flare class. The C II and Mg II lines show a similar tendency but with smaller velocities compared to the Si IV line. Additionally, the Mg II h or k line shows a blue-wing enhancement in the three flares in particular at the flare ribbon fronts, which is supposed to be caused by an upflow in the upper chromosphere due to the heating of the atmosphere. Moreover, the Mg II h or k line exhibits a central reversal at the flare ribbons, but turns to pure emission shortly after 1--4 minutes. Correspondingly, the C II line also shows a central reversal but in a smaller region. However, for the Si IV line, the central reversal is only found in the X-class flare, but not in the other two flares. As usual, the central reversal of these lines can be caused by the opacity effect. This implies that in addition to the optically thick lines (C II and Mg II lines), the Si IV line can become optically thick in a strong flare, which is likely related to the nonthermal electron beam heating.

\end{abstract} 

\keywords{Solar activity (1475); Solar flares (1496); Solar transition region (1532); Solar chromosphere
(1479); Solar ultraviolet emission (1533); Solar flare spectra (1982)}

\section{Introduction} \label{sec: introduction}

It is known that solar eruptions, such as solar flares and coronal mass ejections, are powered by magnetic reconnection in the solar atmosphere, which can release large amounts of magnetic energy in a short time. In the standard flare model \citep{CSHKP_C_1964,CSHKP_S_1966,CSHKP_H_1974,CSHKP_KP_1976}, the released energy causes plasma heating, particle acceleration, and mass motions at the reconnection site. It is then transported to the chromosphere along the newly reconnected magnetic filed lines through thermal conduction and/or particle bombardment. The deposited energy in the chromosphere causes a local heating there and produces enhanced radiations in chromospheric lines, which are seen as the flare ribbons. At the same time, the local thermal pressure drives the heated plasma upward to the corona along the newly formed flare loops, which is termed as chromospheric evaporation \citep{Neupert_1968,1974_Hirayama_Model_of_evap,1982_Acton_obs_of_evap}. Due to the conservation of momentum, the evaporation process is accompanied by a downward moving condensed plasma, which is called chromospheric condensation \citep{Fisher_1987_condensation,1990_Canfield_momentum_banlance}. In the electron-beam-driven evaporation model, the thermal emission in soft X-rays (SXRs) from the evaporated hot plasma is roughly proportional to the time integral of the nonthermal emission in hard X-rays (HXRs) emitted by nonthermal electron beams, which is known as the Neupert effect \citep{Neupert_1968}. This Neupert effect is also invoked when studying the mechanisms of energy transport during solar flares \citep[][and references therein]{1999_McTiernan_Neupert_effect_evap,Liu_2006_HXR_evap_Neupert,2022_Veronig_Neupert_statics}, especially for large and impulsive flares.

The different kinds of mass motions in the atmosphere during solar flares can be detected by the Doppler shifts of spectral lines formed at different heights. There have been many spectroscopic observations that can be devoted to the study on the chromospheric evaporation and condensation processes. The former is usually shown as blueshifts or blueshifted components of the spectral lines in SXR and extreme-ultraviolet (EUV) passbands \citep{Neupert_1968}, while the latter is mostly evidenced by redshifts or redshifted components of the spectral lines in UV and optical passbands. According to the patterns of Doppler shifts in different spectral lines, there are two types of chromospheirc evaporation, explosive evaporation and gentle evaporation. When both upward motions of hotter plasma and downward motions of cooler plasma can be observed, this process is called explosive evaporation. Previous spectroscopic studies have reported blueshift velocities of hot coronal lines (e.g., Ca XIX and Fe XVI to Fe XXIV lines) from $\sim$50 to $\sim$300 km s$^{-1}$ \citep{1980_Doschek_Ca,1996_Ding,Brosius_2004_EUV_of_evaporation,Milligan_2006_explosive_evap,2006_CDS_Fe_XIX,2010_Chen_Ding_explosive_evap,Doschek_2013_EIX_Fe,Tian_2014_evap,Polito_2015_Blueshifts,Tian_2018_evaporation,2019_Dong_precursor_nonthermal} and redshift velocities of cool lines formed in the chromosphere to the transition region (TR; e.g., H$\alpha$, Mg II h\&k, C II, and Si IV lines) from $\sim$15 to $\sim$80 km s$^{-1}$ \citep{1994_Ha_conde,1995_Ding_Ha_model_conden,2005_Kamio_CDS,Li_Ying_2011_EIS,2015_Milligan_review_spec_flares,Graham_2015,2020_Yu,2022_Li_Dong_Sim_obs_explos}. On the other hand, when only upward motions are detected in spectral lines with almost no downward motions appearing, the evaporation is referred to as gentle evaporation. In events of gentle evaporation, the heated plasma has a relatively smaller blueshift velocity, such as tens of km s$^{-1}$ \citep{Milligan_2006_gentle_evaporation,Brosius_2009_explosive_to_gentle_evap,2009_obs_EBTEL_gentle_explosive,Li_Ying_2011_EIS,Sadykov_2015,Li_Ying_2015_gentle,2019_Ying}.

During solar flares, the Si IV resonance lines at 1393.755 and 1402.770 \AA\ can provide spectroscopic diagnostics of the transition region structure and dynamics. In most previous studies, the Si IV resonance lines were considered to be optically thin, whose profiles fit the Gaussian shape very well. Theoretically, the intensity ratio of these two lines is equal to 2 under the optically thin condition, which applies well to the quiet-Sun region \citep{1999_Mathioudakis_ratio_theory,2020_Tripathi_Ratio_distribution}. In the optically thick case, the intensity ratio could deviate from 2. Due to the opacity effect, the line ratio usually falls below 2 \citep{2018_Del_review_spectral}. For example, \citet{2018_Gontikakis_resonant_scattering} studied the characteristics of the line ratio in an active region (AR), and found that it can be smaller than 2, which should be ascribed to the opacity effect since the line profiles are self-reversed. \cite{Kerr_2019} simulated the Si IV resonance lines under different electron-beam heating cases during flares, and found that the opacity effect becomes more important in those cases when the column density of the line forming region increases sufficiently, which is related to the electron beam parameters, such as the total energy, low-energy cutoff, and spectral index. On the other hand, the line intensity ratio could also be greater than 2 if the resonant scattering is dominant \citep{2018_Gontikakis_resonant_scattering}. More interestingly, in solar flares cases, the line intensity ratio varies significantly with wavelength across the line. Such a wavelength-dependent intensity ratio can be smaller than 2 in the line core but greater than 2 in line wings; however, the ratio of the integrated intensity ratio is still roughly equal to 2, as reported by \citet{2022_Zhou_opacity}. The reason is that the line core becomes optically thick while the line wings remain optically thin.

It is known that flare ribbons represent the footpoints of post-reconnected flare loops. The front or the leading edge of flare ribbons is of special interest since it maps the newly reconnected flare loops. In the past, much attention has been paid to the spectral features at the leading edge. \cite{2016_Xu_He_dimming,2022_Xu_He_dimming} observed an enhanced absorption of the He I 10830 \AA\ line at the leading edge of a flare ribbon, as well as obvious central reversals in the Mg II and C II lines. To explore the formation mechanisms of the He I 10830 \AA\ line, \cite{2021_Kerr_absorp_He_I} conducted radiative hydrodynamic simulations of heated model atmospheres of flares. They confirmed that collisional ionization from the nonthermal electron beam followed by reconnection, as proposed by \cite{2005_Ding_He_I}, is the chief mechanism to reproduce the enhanced absorption, the extent of which is correlated with the low-energy cutoff of the electron beam. \cite{2021_Panos_2021} found that in the flare ribbon front, not only the Mg II and C II lines show central reversals, but also the Si IV line has an obvious dip near the line core. In addition, the central reversal of Mg II line can turn to a single emission peak within 1--3 minutes, which is caused by a sufficient heating of the chromosphere. \cite{2023_Polito_front} quantitatively analyzed the spectral characteristics of Mg II k and triplet lines at flare ribbon fronts, and also performed numerical simulations of different heating models for the leading edge and the trailing parts of the ribbons. They found that at the flare ribbon front, the nonthermal electron beam has a larger low-energy cutoff with, however, a more modest and more gradual energy flux than the trailing region, which implies that the energy there is deposited deeper although weaker.

In this paper, we study the spectral characteristics of Si IV, C II, and Mg II lines for three solar flares. We extensively catalogue the behavior of the observed lines, which could be employed for inferring the the energy transport processes and heating mechanisms in these events. In Section \ref{sec: data reduction}, we introduce the observational data and the analysis methods. Section \ref{sec: results} presents a brief description of the three flares as well as the detailed results. Finally, we summarize and discuss the results in Section \ref{sec: sum and discuss}.

\section{Observational data and analysis methods} \label{sec: data reduction}

\subsection{Observational Data} \label{sec: obs}

The three flares under study were well observed by the Interface Region Imaging Spectrograph \citep[IRIS; ][]{2014_IRIS} and a number of space instruments that provide multi-wavelength imaging and spectral observations. The SXR flux class, start time, and peak time of the three flares, together with the instrumental parameters including the cadence and spatial pixel size of the IRIS spectrograph (SG) and the HXR energy band that has a response during the flares, are shown in Table \ref{tab:flare list}.

\begin{center}
\begin{table}[htb]
    %\tablewidth{0pt}
    \centering
    \caption{List of the Flares Analyzed in the Study.}
    \begin{tabular}{c c c c c c c} 
    \hline \hline
    GOES Class & Date & Start Time (UT) & Peak Time (UT) & SG Cadence (s) & Pixel Size ($^{\prime\prime}$) & HXR emission\\
    \hline
    X1.3 & 2022 Mar 30 & 17:21 & 17:37 & 9.6 & 0.166 & STIX 4--84 keV\\
    M1.5 & 2022 Jan 18 & 17:01 & 17:44 & 0.8 & 0.333 & STIX 4--25 keV\\
    C1.6 & 2015 Dec 19 & 10:40 & 10:51 & 3.0 & 0.333 & RHESSI 3--50 keV\\
    \hline
    \end{tabular}
    \label{tab:flare list}
\end{table}
\end{center}

IRIS includes two main parts, SG and slit-jaw imager (SJI), which can provide high resolution spectra and images with different field-of-views (FOVs). The spectrograph slit has a width of 0.3$^{\prime\prime}$ and a length of no longer than 176$^{\prime\prime}$. The three flares were all observed in the sit-and-stare mode with different spatial resolutions and cadences. The spectra cover two far-ultraviolet wavebands (FUV1, 1332--1358 \AA\ and FUV2, 1389--1404 \AA) and a near-ultraviolet waveband (NUV, 2783--2834 \AA), which contain some important lines, such as the Fe XXI 1354.08 \AA\ ($\sim${10$^{7.0}$} K), Si IV 1393.76 and 1402.77 \AA\ ($\sim${10$^{4.8}$} K), C II 1334.53 and 1335.71 \AA\ ($\sim${10$^{4.3}$} K), and Mg II h\&k lines ($\sim${10$^{4.0}$} K). The spectral resolutions are 25.96, 25.44, and 25.46 m\AA\ per pixel for FUV1, FUV2, and NUV wavebands, respectively. Note that the Mg II k spectra are two-pixel binned in the wavelength for the M- and C-class flares. The SJIs are acquired at wavelengths of 1330 and 1400 \AA\ with a bandwidth of 40 \AA\ and 2796 and 2832 \AA\ with a bandwidth of 4 \AA.

We focus on analyzing the data of Si IV 1402.77 \AA, C II 1334.53 \AA, and Mg II h or k lines. Note that the Si IV 1393.76 \AA\ line and sometimes the Mg II h or k line were unavailable in the three flare observations. In addition, we also analyze the coronal Fe XXI line if the data are available. For wavelength calibrations, the theoretical line-center wavelengths are used for these lines. We have tested that the observed line-center wavelengths obtained by averaging the line profiles in a quiet region near the flares have no significant deviation from the theoretical line-center wavelengths, similarly to what has been revealed by \cite{2018_Brosius} and \cite{2019_Ying}.

We also use the 94 \AA\ ($\sim${10$^{6.8}$} K) images observed by the Atmospheric Imaging Assembly \citep[AIA; ][]{2012_AIA} and the line-of-sight (LOS) magnetograms by the Helioseismic and Magnetic Imager \citep[HMI; ][]{2012_HMI}, both of which are on board the Solar Dynamics Observatory \citep[SDO; ][]{2012_SDO}. The full-disk images from AIA have a pixel size of 0.6$^{\prime\prime}$ with a cadence of 12 s for seven EUV bands (94, 131, 171, 193, 211, 304, 335 \AA), 24 s for two UV bands (1600 and 1700 \AA). The magnetograms observed by HMI have a pixel size of 0.5$^{\prime\prime}$ with a cadence of 45 s. The coalignment between SDO/AIA images, HMI magnetograms, and IRIS images can be done automatically in the SolarSoftWare (SSW).

The SXR data are from the Geostationary Operational Environmental Satellites (GOES), which has two wavebands, 0.5--4 and 1--8 \AA. The 1--8 \AA\ flux is used to monitor the response of flares. For HXR observations, the X1.3 and M1.5 flares were captured by the Solar Spectrometer Telescope for Imaging X-rays \citep[STIX; ][]{2020_STIX} on board Solar Orbiter \citep{2020_SOlar_Orbiter_Sci}, and the C1.6 flare was observed by the Reuven Ramaty High Energy Solar Spectroscopic Imager \citep[RHESSI; ][]{2002_RHESSI}. The energy range, energy resolution, and time resolution are 4--150 keV, 1 keV, and 0.1--1 s for STIX, respectivley, while they are 3--1700 keV, $\leq$ 1 keV, and 2 s for RHESSI, respectively. The finest angular resolution is 7$^{\prime\prime}$ and 2.3$^{\prime\prime}$ for STIX and RHESSI, respectively.

\subsection{Analysis Methods} \label{sec: methods}

For the Si IV, C II, and Mg II h or k lines, we first make a moment analysis to calculate the total intensity (the zeroth order moment of the line profiles), line center wavelength (the first order moment), and line width (the second order moment), which are described in detail in \citet{2020_Yu}. In the rest of the paper, we use the units of Doppler velocity instead of that wavelength to describe a line, which are related as $v = c(\lambda_o - \lambda_r)/\lambda_r$, where $c$ is the speed of light, $\lambda_o$ is the observed wavelength, and $\lambda_r$ is the reference wavelength of the line center. We need to mention that the moment method can quickly give a unique result, but the Doppler velocity tends to be underestimated since it only considers the centroid position, especially for the Si IV line with complex profiles, which are often multi-peaked. Therefore, we also use the method of double-Gaussian fitting and the bisector method to derive the Doppler velocity for the Si IV line. As for the Fe XXI line, although it is optically thin, its profile is contaminated by the C I 1354.29 \AA\ line and some other cool lines from neutral and singly ionized atoms. We thus apply a multi-Gaussian fitting to the contaminated Fe XXI line profile in order to obtain the peak intensity, Doppler shift, and line width for the pure Fe XXI line emission.

In addition, we quantitatively evaluate the central reversals of the Si IV, C II, and Mg II line profiles in the flare region. In order to identify the central reversal, we make a derivative of the line profile with respect to wavelength and search for the part that satisfies the condition of $f^{'}(\lambda_1) < 0$ and $f^{'}(\lambda_2) > 0$, where $\lambda_{1}$ and $\lambda_{2}$ are shorter and longer wavelength points, separated by at least 2 wavelength pixels. However, doing so can find out all dips appearing in the profile, some of which are not related to the central reversal but caused by the blueshifted or redshifted components when flaring atmosphere contains significant mass motions like the chromospheric evaporation and condensation. Thus we put a further constraint by requiring that the central position of the dip is within $\pm$2 wavelength pixels from the reference line center (corresponding to about $\pm$11 km s$^{-1}$ for each of the lines). Note that when doing this, we make the Mg II spectra have the same two-pixel binning for all the three flares. We then calculate the relative depth ($R_d$) of the central reversal by the following formula:
\begin{equation}
    R_d = 1 - \frac{2I_c}{I_b + I_r}\label{eq:R_d}
\end{equation}
where $I_c$, $I_b$, and $I_r$ are the intensity of the central reversal, its nearby blue peak intensity, and the red peak intensity, respectively. Generally speaking, a larger value of $R_d$ implies a more significant central reversal. However, the value of $R_d$ can vary with the asymmetries of the line (different intensities of the blue and red peaks); thus it should be used with caution.

\section{Events and Results} \label{sec: results}

\subsection{The X1.3 Flare} \label{subsec:flare 1}

\subsubsection{Event Overview} \label{subsec: overview 1}

\begin{figure}		
    \epsscale{1.15}
    \plotone{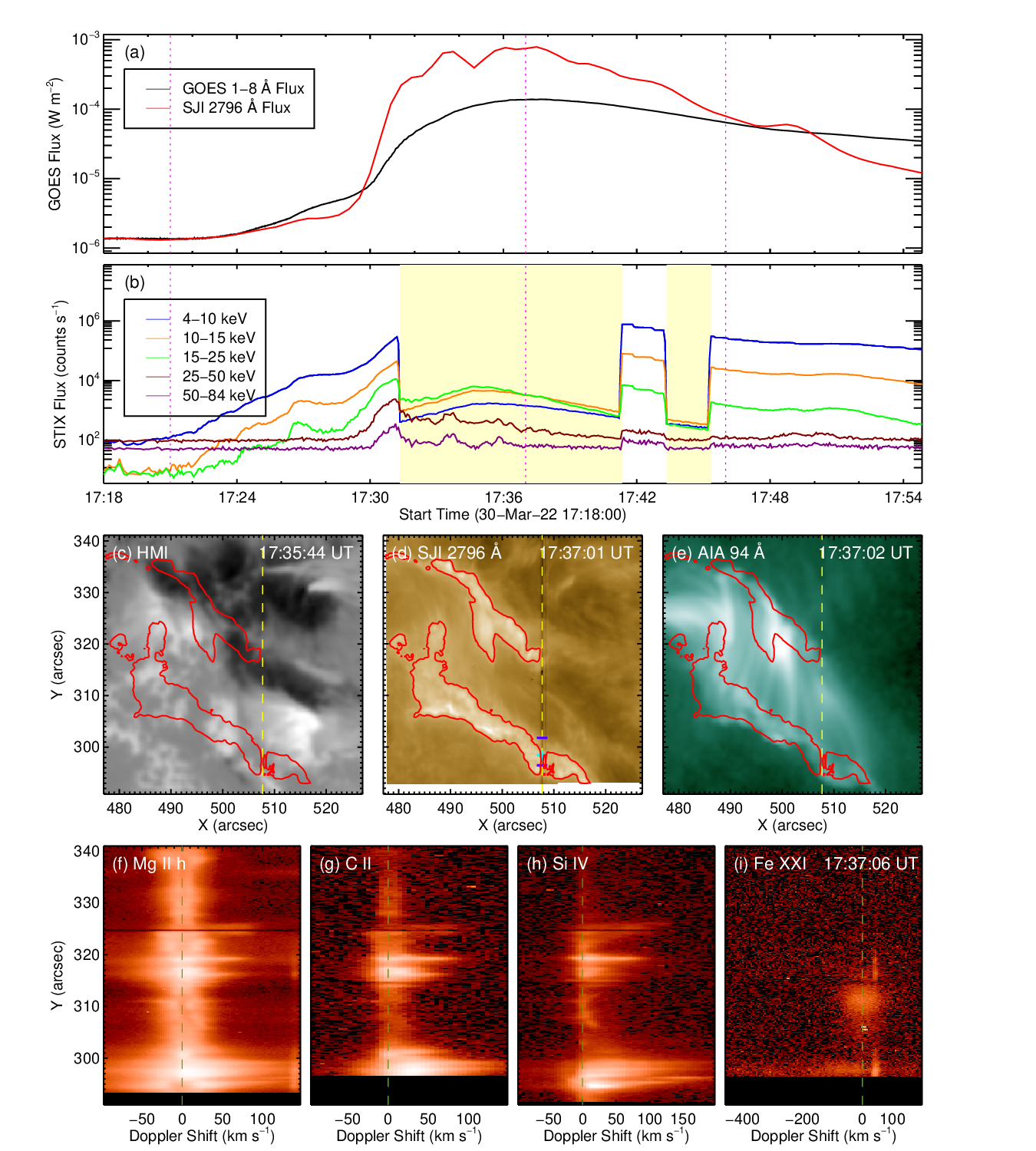}
    \caption{Observation overview of the X1.3 flare on 2022 March 30. (a): GOES 1--8 \AA\ flux and SJI 2796 \AA\ flux (integrated over the flare region as shown in panel (d)). (b): STIX fluxes at different energy bands. The yellow shaded area marks the time period when the attenuator works. The vertical magenta dotted lines in panels (a) and (b) indicate the start, peak, and end times of the flare. (c)--(e): HMI LOS magnetogram, SJI 2796 \AA, and AIA 94 \AA\ images around the flare peak time. The red contours mark the flare ribbons as seen in the SJI 2796 \AA\ image with a level of 15$\%$ of the maximum intensity. The vertical yellow dashed line indicates the IRIS slit position. The two blue bars in panel (d) mark the region used to integrate the Fe XXI spectra in Figure \ref{fig:Fe_time-sequnce_1}(a), and the cyan cross indicates the selected position where the line profiles are shown in Figures \ref{fig:Fe_time-sequnce_1}(c) and (d). (f)--(i): The spectra of Mg II h, C II, Si IV, and Fe XXI along the slit at 17:37:06 UT (i.e., the flare peak time). The olive dashed lines mark the reference wavelengths of these spectral lines.}
    \label{fig:SXR_SJI_1}
\end{figure}

The X1.3 flare, occurring in NOAA AR 12975, started at 17:21 UT, peaked at 17:37 UT, and ended at 17:46 UT on 2022 March 30. IRIS provides spectral data with a cadence of 9.6 s and SJIs at 1330, 1400, and 2796 \AA\, with an FOV of 60$^{\prime\prime}$ $\times$ 65$^{\prime\prime}$ and a cadence of 28 s. The GOES 1--8 \AA\ flux, SJI 2796 \AA\ flux of the flaring region, STIX 4--84 keV fluxes, HMI LOS magnetogram, SJI 2796 \AA\ image, AIA 94 \AA\ image, and the spectra along IRIS slit are shown in Figure \ref{fig:SXR_SJI_1}. The yellow shaded area in Figure \ref{fig:SXR_SJI_1}(b) indicates the time period during which the STIX attenuator works. From the lightcurves of SXR 1--8 \AA, SJI 2796 \AA, and STIX 4--25 keV, one can see that there is a gentle heating before $\sim$17:29 UT, probably as a precursor of the flare. After that, an impulsive heating begins and lasts to the flare peak, accompanied by an enhancement of the HXR emission above 25 keV.

From the SJI at 2796 \AA\ in Figure \ref{fig:SXR_SJI_1}(d), we delineate a pair of roughly parallel ribbons located in two regions with opposite magnetic polarities (see Figure \ref{fig:SXR_SJI_1}(c)). Combined with the image of AIA 94 \AA\ (Figure \ref{fig:SXR_SJI_1}(e)), one can clearly see flare loops filled with hot plasma connecting the two ribbons. The IRIS slit (indicated by the dashed yellow line in Figures \ref{fig:SXR_SJI_1}(c)--(e)) crosses the southern ribbon, part of the northern ribbon, and part of flare loops in between. Note that IRIS observations missed part of the data. Specifically, the data of FUV1 bandpass (C II and Fe XXI lines) at Y $\le$ 296.4$^{\prime\prime}$ and the NUV data (Mg II h\&k lines) at Y $\le$ 293.12$^{\prime\prime}$ are unavailable, as shown in the spectra along the slit around the peak time in Figures \ref{fig:SXR_SJI_1}(f)--(i). The slit spectra show that within the flare ribbon (Y $=$ $\sim$298$^{\prime\prime}$), there appears an obvious enhancement in the red wings of the Mg II, C II, and Si IV lines formed in the chromosphere to the TR, while the coronal Fe XXI line shows a blueshifted enhancement.

\subsubsection{Doppler shifts of the spectral lines} \label{subsec: condensation 1}

\begin{figure}
    \centering
    %\epsscale{0.4}
    \includegraphics[width=0.9\linewidth]{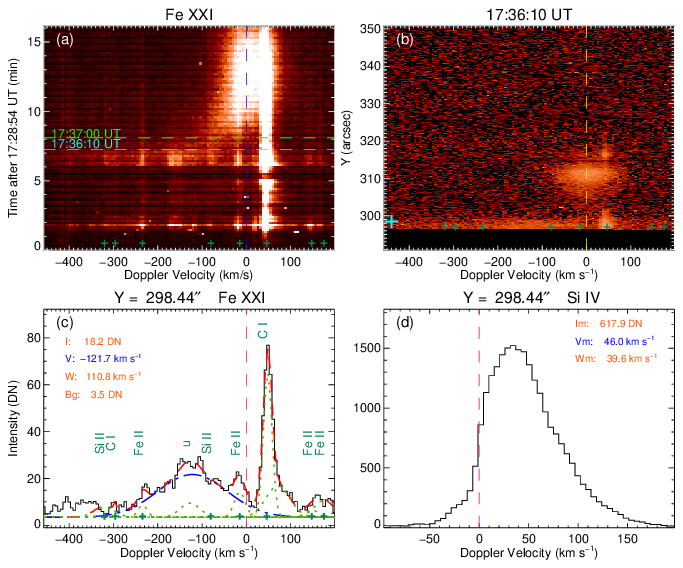} 
    \caption{(a): Temporal evolution of the Fe XXI spectra integrated over the region outlined by two blue bars in Figure \ref{fig:SXR_SJI_1}(d). The vertical blue dashed line indicates the reference wavelength of the Fe XXI line, so do the vertical dashed lines in panels (b) and (c). The dark green crosses (also in panels (b) and (c)) mark the cooler lines in the Fe XXI spectral window. The horizontal dashed lines mark the flare peak time (green) and the time (cyan) of the selected spectra in panels (b)--(d). (b): Fe XXI spectra along the slit at 17:36:10 UT. The cyan cross indicates the selected position of Y $=$ 298.44$^{\prime\prime}$. (c): The observed line profile (black) at the position indicated by the cyan cross in panel (b) and its multi-Guassian fitting (red) including the components of the Fe XXI (blue) and some cooler lines plus an unidentified line (olive). The peak intensity, Doppler velocity, and line width of the Fe XXI component and the background intensity are shown in the panel. (d): The observed Si IV line profile at the same position as the Fe XXI line profile in panel (c). Its total intensity, Doppler velocity, and line width obtained by the moment method are shown in the panel. The vertical pink dashed line indicates the reference wavelength of the Si IV line.}
    \label{fig:Fe_time-sequnce_1}
\end{figure}

Figure \ref{fig:Fe_time-sequnce_1}(a) shows the wavelength-time plot of the Fe XXI spectra integrated over the flare region marked by the two blue bars in Figure \ref{fig:SXR_SJI_1}(d). Note that the exposure time during this observation period was changing, and thus we need to make a normalization for the spectra at each exposure. Within the southern flare ribbon, the Fe XXI spectra start to show a blueshifted enhancement at $\sim$17:35 UT, with a maximum velocity greater than 100 km s$^{-1}$, for example, the spectra at 17:36:10 UT in Figures \ref{fig:Fe_time-sequnce_1}(b) and (c). This blueshifted enhancement is supposed to be related to the chromospheric evaporation at the flare loop footpoints, as in the events studied by \cite{Tian_2018_evaporation}. After about three minutes, the blueshift velocity rapidly decreases to almost zero. It should be mentioned that in the Fe XXI spectral window, there exist a number of cooler lines as indicated by the dark green plus signs in Figures \ref{fig:Fe_time-sequnce_1}(a)--(c). In Figure \ref{fig:Fe_time-sequnce_1}(c), we show an example line profile at the position indicated by the cyan cross in Figure \ref{fig:Fe_time-sequnce_1}(b), as well as the multi-Gaussian fitting results, including the Fe XXI component (blue line) and some cooler and unidentified lines (olive dashed lines). It is seen that the Doppler velocity of the Fe XXI blueshifted component reaches about 121.7 km s$^{-1}$. The corresponding Si IV line profile is shown in Figure \ref{fig:Fe_time-sequnce_1}(d), based on which the moment analysis yields a redshift velocity of 46.0 km s$^{-1}$. The simultaneous detection of blueshifts in the coronal line and redshifts in the transition region line, indicates that explosive chromospheric evaporation occurs in this X1.3 flare, in particular during the impulsive phase \citep{Fisher_1985,Miligan_2006}.

To study in more detail the evolution of the flare, we plot the spatio-temporal distribution of the total intensity, Doppler velocity, and line width for the Si IV and Mg II h lines in Figure \ref{fig:momet_all_1}. The Doppler velocity is derived from the moment method. Note that for the Si IV line, we discard the data points where the total intensity is below a threshold and thus the signal-to-noise ratio (SNR) is too low to yield physically meaningful results (the same thing is done for the following two flares). In Figure \ref{fig:momet_all_1} we also overplot two levels of intensity contours, which are used to define the flare region and the main ribbon of the flare. The maroon contour with a relatively lower level defines the flare region, while the black contour with a higher level refers to the core region of the flare ribbon or the main ribbon. Note that the lower and higher levels of the contours are the same for a certain line in the three flares under study, which are 2\% and 18\% of the peak total intensity for Si IV, 3\% and 11\% for C II, and 18\% and 35\% for Mg II, respectively. We also define the area in the front of the main ribbon as the flare leading edge of the ribbon (or ribbon front), and the area behind it as the trailing region of the ribbon, the later of which is sometimes overlapped with flare loop legs. Within the main ribbon of the X1.3 flare, the Si IV and Mg II h lines exhibit redshifts with average Doppler velocities of 56.6 and 30.9 km s$^{-1}$, respectively. From Figure \ref{fig:momet_all_1} it can also be seen that the line intensity enhancement is spatially and temporally correlated with a larger Doppler velocity and a larger line width generally, which is a typical feature of chromospheric condensation. 

From the spatio-temporal maps, one can find that the south region of the flare repetitively brightens from 17:25 UT and each lasts about one minute before the main ribbon appears at Y $=$ $\sim$301.5$^{\prime\prime}$. Such a brightening has a spatial size of smaller than 0.5$^{\prime\prime}$ (see also Figure \ref{fig:dip_1}). It is interesting that such a small-scale repeated brightening is associated with a variable redshift velocity, somewhat similar to the intensity and velocity fluctuations in the chromospheric and TR lines observed by \cite{Brosius_2015} and \cite{Warren_2016_extended_impulsive_heating} and simulated by \cite{Reep_2016_simulation} in the scenario of multi-threaded loops. However, unlike that in the previous studies, the brightenings here seem to appear as the precursor (17:25--17:29 UT) before the impulsive phase of the flare. Overall the heating is gentle in this precursor phase (see also Figure \ref{fig:SXR_SJI_1}(a)), when the HXR emission has a response only below 25 keV (see Figure \ref{fig:SXR_SJI_1}(b)). It should also be mentioned that the repeated brightenings are located near a sunspot penumbra and the line intensity and redshift velocity might be modulated by the penumbral waves. After 17:29 UT, the brightening in the southern part of the flare region begins to move southward along the slit, while the northern brightening moves away from the southern one after about 17:36 UT. Both brightening regions are associated with redshifted spectral lines of Si IV, C II, and Mg II. Within the main ribbon (with a spatial size of 0.5--3$^{\prime\prime}$), the redshifts last about 1--3 minutes, consistent with the typical duration of chromospheric condensation \citep{Fisher_1989,Ding_1995_velocity_filed}. Outside the main flare ribbon, there is a bright region at Y $=$ $\sim$299$^{\prime\prime}$ at $\sim$17:35 UT, in which the Si IV line shows a redshift velocity of over 100 km s$^{-1}$. Since this region lies in the inner part of the flare region, it is likely overlapped by the loop legs with the footpoints. Thus, the high speed redshift happens when cooled plasma falls down from the loop top. Besides, there appears a long lived redshift at the inner part of the flare region untill $\sim$17:52 UT, and the redshift also exists in the area between the two parts of the flare region from $\sim$17:46 UT, which will be discussed in Section \ref{sec: sum and discuss}. At $\sim$17:36 UT, the Mg II h line at the leading edge of the flare ribbon shows some blueshifts or blue-wing enhancements (see Figure \ref{fig:momet_all_1}(d)), which will also be discussed in Section \ref{sec: sum and discuss}. 

\subsubsection{Central reversals of the spectral line profiles} \label{subsec: central reversal 1}

\begin{figure}		
    \centering
    \plotone{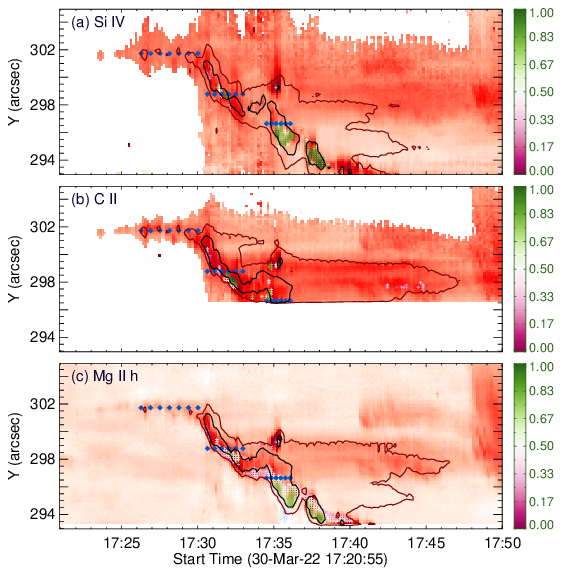}
    \caption{Spatio-temporal distributions (colored plus signs) of the central reversal of the Si IV, C II, and Mg II h line profiles at the southern part of the flare region overplotted on the Doppler velocity maps. The relative depth ($R_d$) is indicated in the color bar. The contours indicate the flare regions with levels of (2$\%$, 18$\%$), (3$\%$, 11$\%$), and (18$\%$, 35$\%$) of the peak total intensities for the Si IV, C II, and Mg II h lines, respectively, with the black contour marking the main ribbon. The colored diamonds indicate the selected positions and time steps for the line profiles as shown in Figure \ref{fig:dip_sample_1}. There is an intensity threshold of 15 DN for the Si IV and C II lines.} 
    \label{fig:dip_1}
\end{figure}

\begin{figure}
    \centering
    \epsscale{1.1}
    \plotone{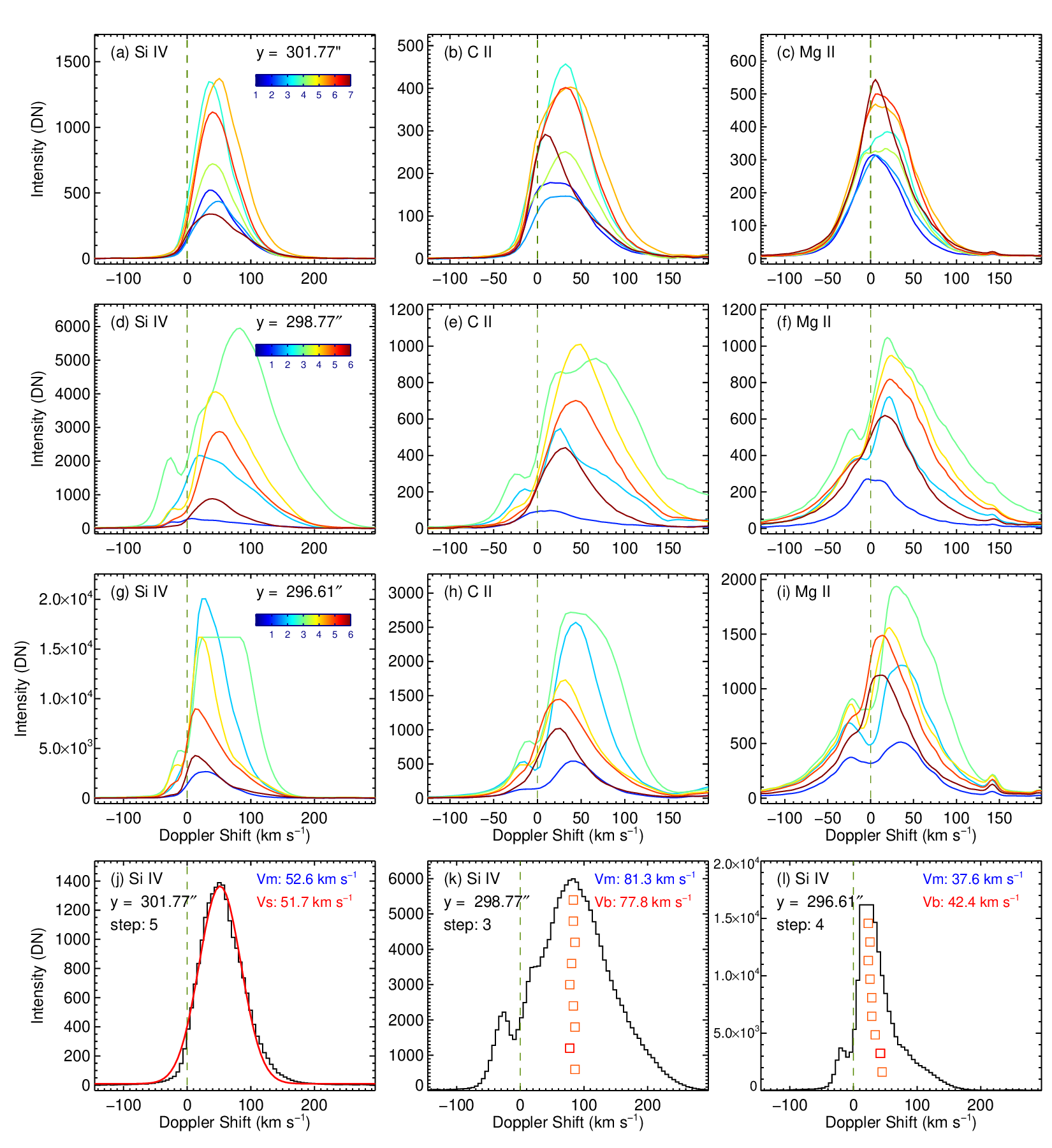}
    \caption{(a)--(i): Si IV, C II, and Mg II line profiles at the selected positions and time steps marked by the colored diamonds in Figure \ref{fig:dip_1}. Different colors represent different time steps. (j)--(l): Si IV line profiles (black) at the three selected positions when the Doppler velocity is relatively larger for the selected time steps. For the Si IV line profile at the first position, we use a single-Gaussian fitting (red) to obtain the Doppler velocity ($V_s$). For the other two profiles, the bisector method is applied to derive the Doppler velocity ($V_b$). The velocities by using different methods (including the one by the moment method, $V_m$) are given in the panels. Note that $V_b$ is the velocity at the 20$\%$ level of the peak intensity. The vertical dashed line in each of the panels denotes the reference wavelength of the spectral line.}
    \label{fig:dip_sample_1}
\end{figure}

The spatio-temporal distributions of the central reversal of Si IV, C II, and Mg II h line profiles for the southern part of the flare region are plotted by colored plus signs over the Doppler velocity maps in Figure \ref{fig:dip_1}, where the reversal magnitude ($R_d$) is shown in the color bar. Note that here we only show the results of the central reversal within the flare region as defined by a lower level of the intensity contour in the above section. Before 17:30 UT, none of the three lines has a central reversal, although the C II and Mg II h lines are usually optically thick. It should be mentioned that the southern ribbon during this period is near a sunspot penumbra, in which the Mg II and C II lines usually show a single-peak profile. After 17:30 UT, the line profiles begin to show a central reversal, especially the Si IV line within the main ribbon part with a larger line intensity, as well as several points at the leading edge of the ribbon. Due to the limitation of the method here, we may miss some line profiles with only a very weak central reversal, such as the profiles plotted as the red lines in Figures \ref{fig:dip_sample_1}(g) and (i). As shown in Figure \ref{fig:dip_1}, for the Si IV and C II lines, the central reversal appears in small-scale patchy regions, while for the Mg II h line, it appears in a much larger region. Specifically, for the Si IV line, the central reversal mainly appears at the main ribbon, and the size of the area is from 0.5$^{\prime\prime}$ to 2$^{\prime\prime}$, and the duration is from 10 to 80 s. The magnitude of central reversal, $R_d$, varies with space and time from 0.02 to 0.89 with an average value of 0.69. As for the C II line, the central reversal appears not only at the main ribbon part, but also at the leading edge of the flare ribbon. The spatial size of the central reversal is about 0.5--1$^{\prime\prime}$ and its duration is about 10--60 s. It should be noted that the spatial size and duration of the central reversal of C II are likely underestimated since the data at Y $\le$ 296.4\arcsec\ were missing. The magnitude of $R_d$ of the C II line profiles ranges from 0.06 to 0.86, with the values from the main ribbon greater than the ones at the leading edge. Unlike the sporadic occurrence of the central reversal in the Si IV and C II lines, the central reversal in the Mg II h line can appear in a larger region of a spatial size of 2.5$^{\prime\prime}$, which can cover the ribbon front, the main ribbon, and also the trailing region and lasts for about 3 minutes. In general, the value of $R_d$ is larger at the main ribbon (with an average of 0.50) than that at the leading edge (smaller than 0.50 for the most).

Figures \ref{fig:dip_sample_1}(a)--(i) plot the time series of line profiles at three selected positions along the slit, which are marked by dark blue diamonds in Figure \ref{fig:dip_1}. Figures \ref{fig:dip_sample_1}(j)--(i) give the results by the moment, Gaussian fitting, or bisector methods of the Si IV line for the three positions at the time step when the line profile is typical (say, showing a large Doppler velocity or being centrally reversed). At the first selected position (Y $=$ 301.77$^{\prime\prime}$) during the precursor phase of the flare, all the three lines are singly peaked and redshifted (Figures \ref{fig:dip_sample_1}(a)--(c)). The maximum Doppler shift calculated from the Si IV line by the moment method reaches 52.6 km s$^{-1}$, and the Gaussian fitting result is 51.7 km s $^{-1}$ (Figure \ref{fig:dip_sample_1}(j)). At the second and third selected positions, the line profiles for these three spectral lines all exhibit a central reversal and show a large red asymmetry (red-wing enhancement) at some time instants in the impulsive phase (Figures \ref{fig:dip_sample_1}(d)--(i)). The redshift velocities of the Si IV line are 81.3 and 37.6 km s$^{-1}$ (the velocities at the 20$\%$ bisector level are 77.8 and 42.4 km s$^{-1}$) for the second and third positions, respectively (Figures \ref{fig:dip_sample_1}(k) and (l)). It is seen that the central reversal and line asymmetry change with the development of the flare. The central reversal usually appears when the line intensity increases, together with a more obvious red asymmetry. Note that the Si IV line could be saturated at the flare peak time (see the line profile in green in Figure \ref{fig:dip_sample_1}(g)). 

\subsection{The M1.5 Flare} \label{subsec:flare 2}

\subsubsection{Event Overview} \label{subsec: overview 2}

\begin{figure}		
    \epsscale{1.15}
    \plotone{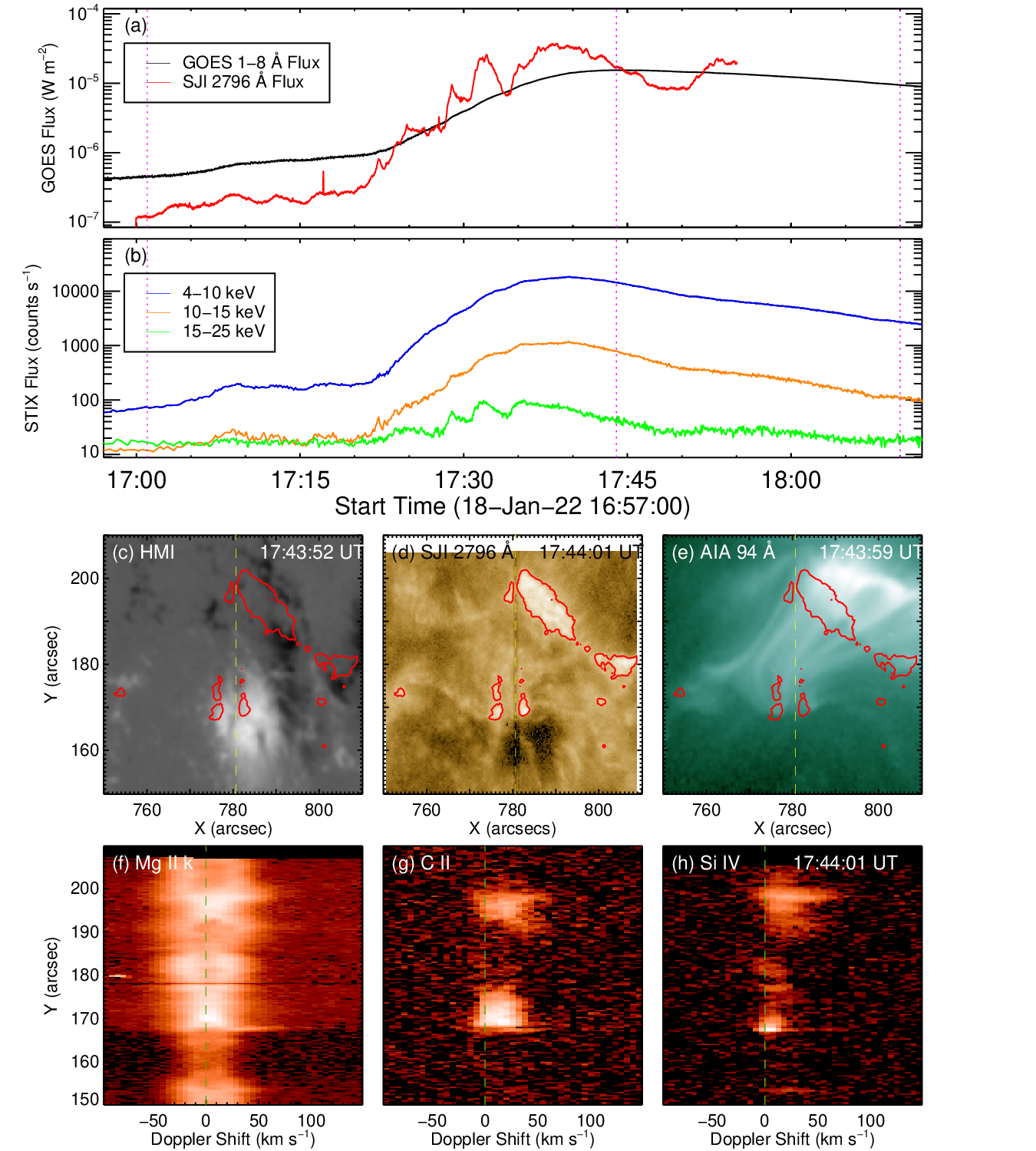}
    \caption{Observation overview of the M1.5 flare on 2022 January 18. (a): GOES 1--8 \AA\ flux and SJI 2796 \AA\ flux (integrated over the flare region as shown in panel (d)). (b): STIX fluxes at different energy bands. The vertical magenta dotted lines in panels (a) and (b) indicate the start, peak, and end times of the flare. (c)--(e): HMI LOS magnetogram, SJI 2796 \AA, and AIA 94 \AA\ images around the flare peak time. The red contours mark the flare ribbons as seen in the SJI 2796 \AA\ image with a level of 40$\%$ of the maximum intensity. The vertical yellow dashed line indicates the IRIS slit position. (f)–(h): The spectra of Mg II k, C II, and Si IV along the slit at 17:44:01 UT (i.e., the flare peak time). The olive dashed lines mark the reference wavelengths of these spectral lines.}
    \label{fig:SXR_SJI_2}
\end{figure}

The M1.5 flare occurred in NOAA AR 12929 on 2022 January 18. It started at 17:01 UT, peaked at 17:44 UT, and ended around 18:10 UT. IRIS acquired spectral data of the flare in the sit-and-stare mode with a cadence of 0.8 s and SJIs at 2796 \AA\ with an FOV of 60$^{\prime\prime}$ $\times$ 62$^{\prime\prime}$ and a cadence of 1 s. Figure \ref{fig:SXR_SJI_2} summarizes the multi-wavelength observations of the flare, including GOES SXR and STIX HXR fluxes, HMI magnetogram, IRIS SJI at 2796 \AA, AIA 94 \AA\ image, and slit spectra of the Mg II k, C II, and Si IV lines around the peak time. One can see that the 2796 \AA\ flux integrated within the flare region has obvious periodic oscillations with a period of 2--3 minutes before the flare peak (Figure \ref{fig:SXR_SJI_2}(a)), roughly corresponding to the oscillations in the HXR 15--25 keV flux (Figure \ref{fig:SXR_SJI_2}(b)). By combining the magnetogram and the 2796 \AA\ and 94 \AA\ images (Figures \ref{fig:SXR_SJI_2}(c)--(e)), one can identify a pair of flare ribbons located in the positive and negative magnetic polarities, connected by the flare loops with hot plasma. It is seen that the IRIS slit (dashed yellow lines) cuts across northern end of the northern ribbon and patchy bright regions of the southern one. As shown in in Figures \ref{fig:SXR_SJI_2}(f)--(h), the Mg II k, C II, and Si IV lines all show red-shifted emissions at the flare ribbons. Note that there was no observation of the Fe XXI line for this flare. This flare has also been analyzed in \citet{2022_M1.5_flare}, focusing on the rapid variations of the Si IV spectra in the southern ribbon. The authors found that the Si IV line in the southern ribbon has a distinct redshifted component caused by strong condensation flows, which corresponds to quasi-periodic pulsations (QPPs) with periods below 10 s in the Si IV line. In addition, \cite{2022_M1.5_flare} found that the GOES flux derivative and the microwave flux also show QPPs, which is proposed to be caused by repeated magnetic reconnection.

\subsubsection{Doppler shifts of the spectral lines} \label{subsec: condensation 2}

\begin{figure}		
    \epsscale{1.25}
    \plotone{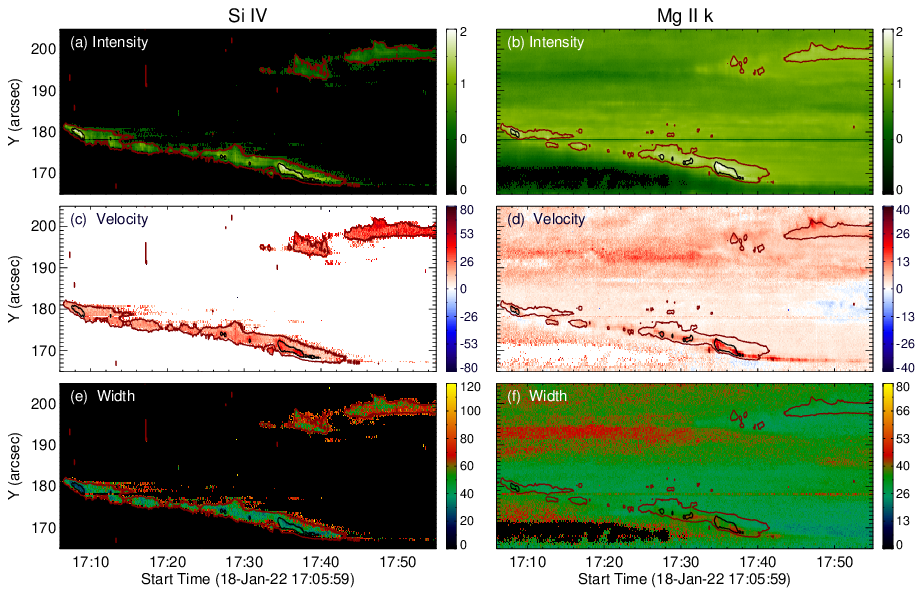}
    \caption{Spatio-temporal variations of the total intensity (in units of log(DN)), Doppler velocity (km s$^{-1}$), and line width (km s$^{-1}$) of the Si IV and Mg II k lines along the IRIS slit obtained from the moment analysis. The contours indicate the flare regions with levels of (2$\%$, 18$\%$) and (18$\%$, 35$\%$) of the peak total intensities for the Si IV and Mg II k lines, respectively, with the black contour marking the main ribbon. Note that for these two lines, we set an intensity threshold of 5 DN for a reliable SNR.}
    \label{fig:momet_all_2}
\end{figure}

The spatio-temporal maps of the total intensity, Doppler shift, and line width of the Si IV and Mg II k lines are plotted in Figure \ref{fig:momet_all_2}. As shown in the line intensity maps (Figures \ref{fig:momet_all_2}(a) and (b)), the brightening region in the southern flare ribbon crossed by the IRIS slit gradually moves towards south. Within the main ribbon (indicated by the black contour), the Si IV and Mg II k lines show redshift velocities with average values of 25.6 and 14.4 km s$^{-1}$, respectively (Figures \ref{fig:momet_all_2}(c) and (d), with spatial size of 1--3$^{\prime\prime}$ and duration of 1--4 minutes). In general, there is a positive correlation between the line intensity and the redshift velocity or line width, especially in the flare main ribbon, which is a typical spectral characteristic of the chromospheric condensation process. However, at some particular points, the Si IV or Mg II line shows a lower line intensity but a larger redshift velocity and a larger line width. These points are mainly located outside the flare region as defined here, where the SNR is relatively lower. One can notice that the Mg II k line shows a blueshift velocity in the front of the main ribbon, say, during 17:08--17:10 UT and 17:36--17:38 UT (Figure \ref{fig:momet_all_2}(d)), similar to the X1.3 flare.

\subsubsection{Central reversals of the spectral line profiles}
\label{subsec: central reversal 2}

\begin{figure}		   
    %\epsscale{0.8}
    \centering
    \plotone{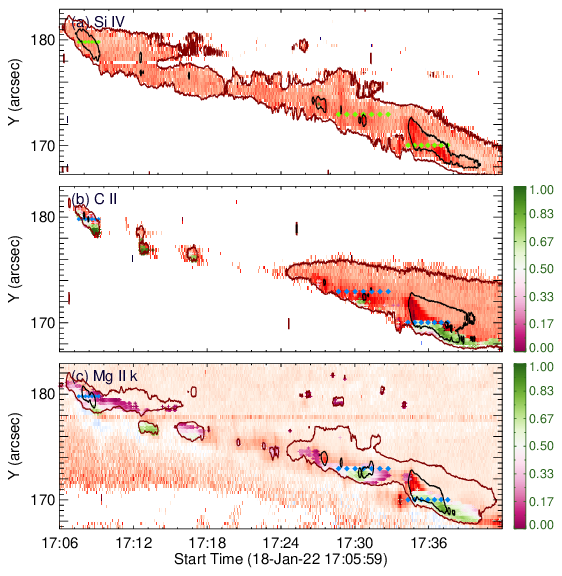}
    \caption{Spatio-temporal distributions (colored plus signs) of the central reversal of the Si IV, C II, and Mg II k line profiles at the southern part of the flare region overplotted on the Doppler velocity maps. The relative depth ($R_d$) is indicated in the color bar. The contours indicate the flare regions with levels of (2$\%$, 18$\%$), (3$\%$, 11$\%$), and (18$\%$, 35$\%$) of the peak total intensities for the Si IV, C II, and Mg II k lines, respectively, with the black contour denoting the main ribbon. The colored diamonds mark the selected positions and time steps for the line profiles as shown in Figure \ref{fig:dip_sample_2}. There is an intensity threshold of 5 DN for the three lines.}
    \label{fig:dip_2}
\end{figure}

Figure \ref{fig:dip_2} shows the spatio-temporal distributions of the central reversal of the Si IV, C II and Mg II k line profiles at the southern part of the flare (the colored plus signs, overplotted on the Doppler velocity maps). It is very interesting that in this M-class flare, there is no obvious central reversal in the Si IV line at the flare ribbon, unlike what appears in the X1.3 flare. The C II line exhibits a central reversal mainly at the leading edge and the core region of the ribbon, especially during the period after $\sim$17:35 UT (about 9 minutes before the flare peak). Note that the central reversal of C II can also appear in the trailing part of the ribbon, say, after $\sim$17:38 UT. The spatial size of the central reversal for the C II line is about 0.5--1$^{\prime\prime}$ and its lifetime is 1--2 minutes. The magnitude of $R_d$ varies from 0.05 to 0.99, with the values from the main ribbon greater than the ones at the leading edge. As for the Mg II k line, the central reversal appears at the ribbon front, the main ribbon, and also the trailing region in this flare. The central reversal of Mg II k appears in a relatively larger spatial size of 0.5--2$^{\prime\prime}$ and lasts for a relatively longer time of 1--4 minutes. The magnitude of $R_d$ is larger at the main ribbon (on average of 0.50), accompanied by a larger redshift velocity, similar to that in the X1.3 flare. Besides, there are some patchy brightening points outside the main flare ribbon (separated contours as seen in Figure \ref{fig:dip_2}(c)), likely corresponding to the footpoints of flare loops as well, in which the Mg II k line also shows a central reversal. 

\begin{figure}
    \centering
    \epsscale{1.1}
    \plotone{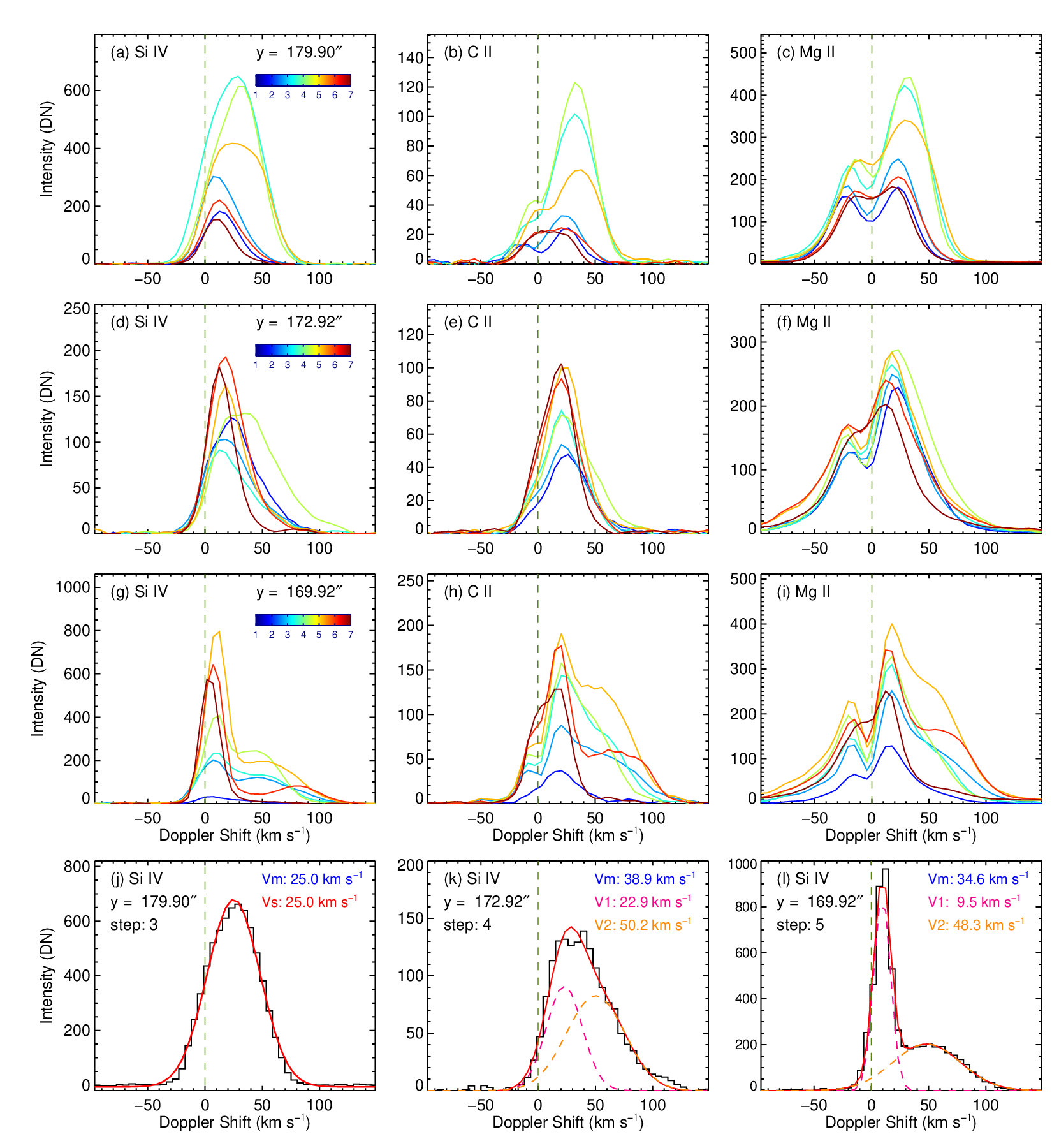}
    \caption{(a)--(i): Si IV, C II, and Mg II line profiles at the selected positions and time steps marked by the colored diamonds in Figure \ref{fig:dip_2}. Different colors represent different time steps. (j)–(l): Si IV line profiles (black) at the three selected positions when the Doppler velocity is the largest among the selected time steps. For the Si IV profile at the first position, we use a single-Gaussian fitting (red) to obtain the Doppler velocity ($V_s$). For the other two profiles, the double-Gaussian fitting (red) is applied to derive the Doppler velocities for different components ($V_1$ and $V_2$, in magenta and orange, respectively). The velocities by using different methods (including the one by the moment method, $V_m$) are given in the panels. The vertical dashed line in each of the panels denotes the reference wavelength of the spectral line.} 
    \label{fig:dip_sample_2}
\end{figure}

The time series of the profiles of the three spectral lines and the Gaussian fitting to the Si IV line at three selected positions (indicated by colored diamonds in Figure \ref{fig:dip_2}) are plotted in Figure \ref{fig:dip_sample_2}. Notably, the Si IV line profiles are mostly singly peaked with a redshift or red-asymmetry but without an obvious central reversal in this M-class flare. In some cases (like the third selected position), the red-shifted component is so prominent that the whole profile looks double peaked, but this does not mean a central reversal. At the first selected position, the Si IV line profile can be well fitted by a single-Gaussian function (Figure \ref{fig:dip_sample_2}(j)). By comparison, those profiles of the Si IV line with an obvious red-asymmetry at the second and third positions can be better fitted with a double-Gaussian function. Figure \ref{fig:dip_sample_2}(l) shows a typical example of the Si IV line profile at the third position. Its fitting comprises a primary component, which is relatively narrower with a smaller redshift velocity (magenta dashed line), and a red-shifted component, which is relatively broader with a larger velocity (orange dashed line). These two components have also been studied in \cite{2022_M1.5_flare}. They were identified to be formed in an underlying stationary region and a condensation region, respectively. Moreover, the C II and Mg II lines at this position also have an obvious redshifted component, together with a central reversal.

\subsection{The C1.6 Flare} \label{subsec:flare 3}

\subsubsection{Event Overview} \label{subsec: overview 3}

\begin{figure}		
    \epsscale{1.15}
    \plotone{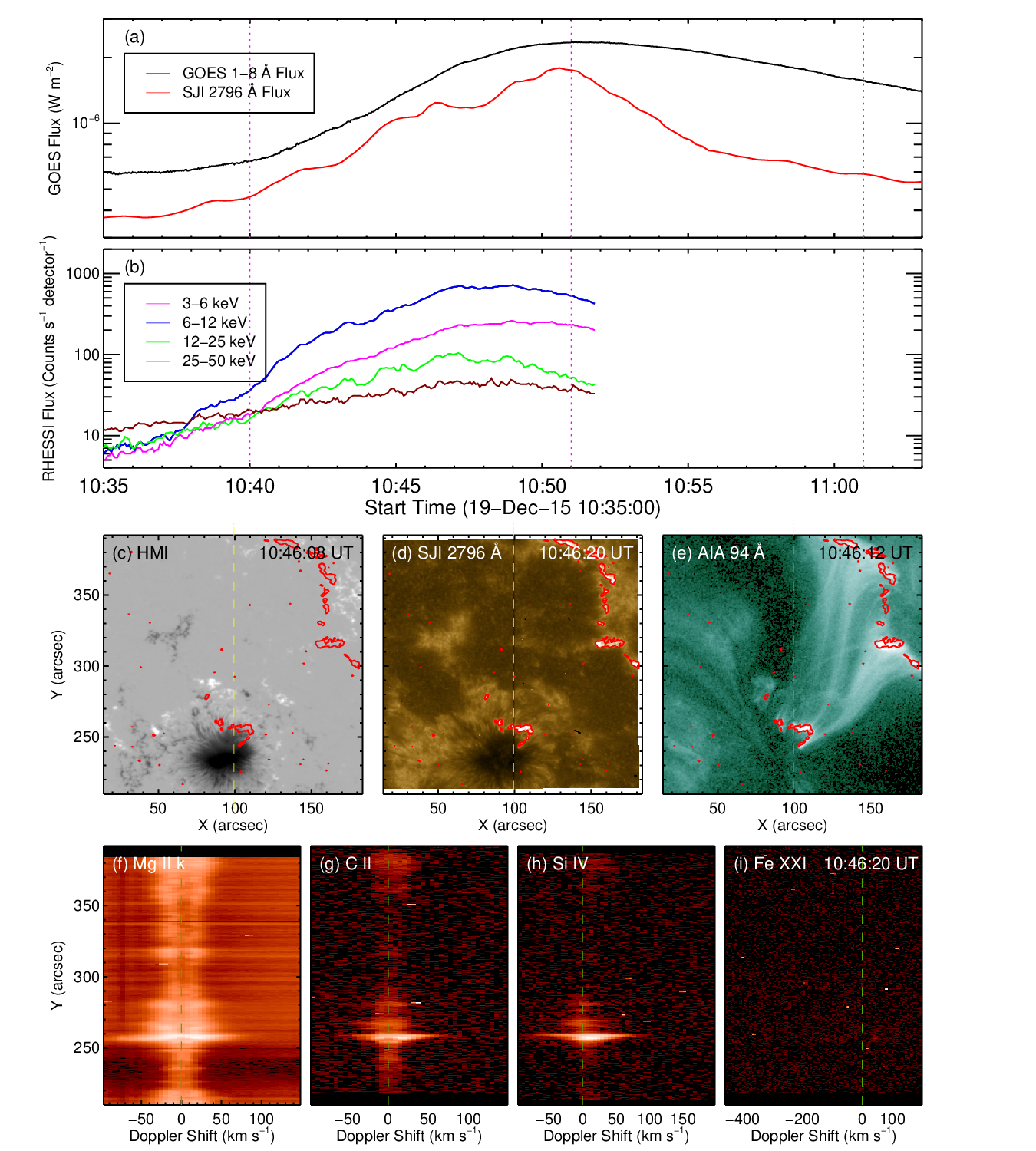}
    \caption{Observation overview of the C1.6 flare on 2015 December 19. (a): GOES 1--8 \AA\ flux and SJI 2796 \AA\ flux (integrated over the flare region as shown in panel (d)). (b): RHESSI fluxes at different energy bands. The vertical magenta dotted lines in panels (a) and (b) indicate the start, peak, and end times of the flare. (c)--(e): HMI LOS magnetogram, SJI 2796 \AA, and AIA 94 \AA\ images around the flare peak time. The red contours mark the flare ribbons as seen in the SJI 2796 \AA\ image with a level of 15$\%$ of the maximum intensity. The vertical yellow dashed line indicates the IRIS slit position. (f)--(i): The spectra of Mg II k, C II, Si IV, and Fe XXI along the slit at 10:46:20 UT. The olive dashed lines mark the reference wavelengths of these spectral lines.}
    \label{fig:SXR_SJI_3}
\end{figure}

The C1.6 flare occured in NOAA AR 12470 on 2015 December 19. This event has been studied by \cite{2019_Ying} on the process of explosive chromospheric evaporation. We plot the multi-wavelength observations of the flare in Figure \ref{fig:SXR_SJI_3}. IRIS provides SJIs at 1330, 1400, 2796, and 2832 \AA\ with an FOV of 167$^{\prime\prime}$ $\times$ 175$^{\prime\prime}$ and a cadence of 13 s. As the GOES 1--8 \AA\ flux shows, this flare started at 10:40 UT, peaked at 10:51 UT, and ended around 11:01 UT (Figure \ref{fig:SXR_SJI_3}(a)). The SJI 2796 \AA\ flux of the flare region shows an oscillating evolution before its peak. The HXR flux observed by RHESSI is most prominent at 6--12 keV (Figure \ref{fig:SXR_SJI_3}(b)), with a peak time before the GOES SXR and SJI 2796\AA\ peaks. From Figures \ref{fig:SXR_SJI_3}(c)--(e), one can see that the flare loops filled with hot plasma show enhanced emission at 94 \AA\ and the footpoints of these flare loops correspond to two bright ribbons that have opposite magnetic polarities. Note that only the southern ribbon was captured by the IRIS slit. The IRIS spectra at NUV and FUV bandpasses are plotted in Figures \ref{fig:SXR_SJI_3}(f)--(i). The Mg II k, C II, and Si IV lines show obvious emissions in the red wings, while the Fe XXI line is hardly affected. As discussed in \cite{2019_Ying}, these red-wing enhancements are typical spectral characteristics of the explosive evaporation.

\subsubsection{Doppler shifts of the spectral lines} \label{subsec: condensation 3}

\begin{figure}	
    \epsscale{1.25}	
    \plotone{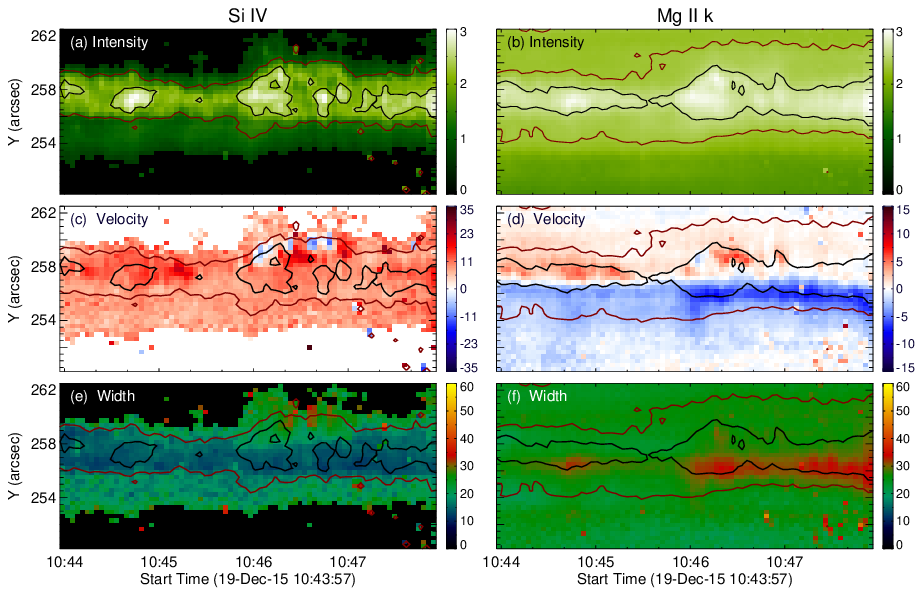}
    \caption{Spatio-temporal variations of the total intensity (in units of log(DN)), Doppler velocity (km s$^{-1}$), and line width (km s$^{-1}$) of the Si IV and Mg II k lines along the slit obtained from the moment analysis. The contours indicate the flare regions with levels of (2$\%$, 18$\%$) and (18$\%$, 35$\%$) of the peak total intensities for the Si IV and Mg II k lines, respectively, with the black contour marking the main ribbon. Note that for the Si IV line, we set an intensity threshold of 5 DN for a reliable SNR.}
    \label{fig:momet_all_3}
\end{figure}

We plot the spatio-temporal evolution of the total line intensity, Doppler velocity, and line width of the Si IV and Mg II k lines at the southern part region during the rise phase of the flare (10:44--10:48 UT) in Figure \ref{fig:momet_all_3}. Beyond this period, the IRIS data are contaminated by compact cosmic rays. It can be seen from the intensity maps that the flare ribbon captured by the IRIS slit keeps brightening in this period. Within the main ribbon (indicated by the black contour), the Si IV line exhibits an average redshift velocity of about 10.5 km s$^{-1}$, lower than that in the former two flares. The Mg II k line, however, shows redshifts with an average velocity of 2.2 km s$^{-1}$ at the northern part of the main ribbon but blueshifts with an average velocity of $-$2.9 km s$^{-1}$ at the southern part, the latter of which corresponds to a broader line width, which is also observed in the former two flares. On the contrary, the Si IV line width at the southern part is smaller than that at the northern part of the main ribbon.

\subsubsection{Central reversals of the spectral line profiles} \label{subsec: central reversal 3}

\begin{figure}		
    %\epsscale{0.8}
    \centering
    \plotone{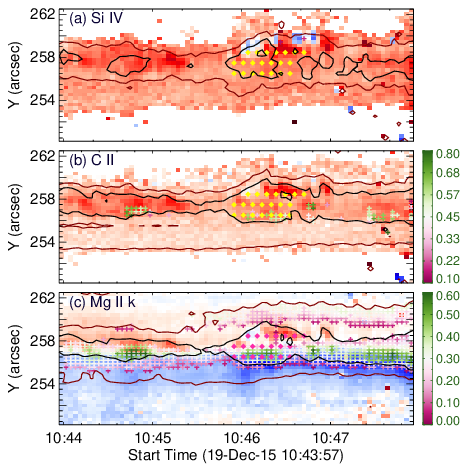}
    \caption{Spatio-temporal distributions (colored plus signs) of the central reversal of the Si IV, C II, and Mg II k line profiles at the southern part of the flare region overplotted on the Doppler velocity maps. The relative depth ($R_d$) is indicated in the color bar. The contours indicate the flare regions with levels of (2$\%$, 18$\%$), (3$\%$, 11$\%$), and (18$\%$, 35$\%$) of the peak total intensities for the Si IV, C II, and Mg II k lines, respectively, with the black contour marking the main ribbon. The colored diamonds indicate the selected positions and time steps for the line profiles as shown in Figure \ref{fig:dip_sample_3}. There is an intensity threshold of 5 DN for the Si IV and C II lines.}
    \label{fig:dip_3}
\end{figure}

The spatial and temporal distributions of the central reversal magnitude ($R_d$) of Si IV, C II, and Mg II k line profiles within the flare region are overplotted on the Doppler velocity maps as crosses with different colors in Figure \ref{fig:dip_3}. From \ref{fig:dip_3}(a), we can see that the Si IV line in this C1.6 flare hardly has a central reversal. As shown in Figure \ref{fig:dip_3}(b), the C II line exhibits central reversals with an average magnitude of 0.48 at discrete locations in the southern part of the main ribbon, whose spatial scales are about 0.33--1$^{\prime\prime}$ and duration is about 30 s. Figure \ref{fig:dip_3}(c) shows that at the southern part of the main ribbon and its leading edge, as well as some parts of the northern trailing, the Mg II k line has a central reversal, which appears in a spatial size of 1.5--3.5$^{\prime\prime}$ and lasts for 2.5--4 minutes. The magnitude of the central reversal, $R_d$, is larger at the main ribbon than at the leading edge, similar to the former two flares. Furthermore, among the centrally reversed Mg II k line profiles at the main ribbon, a larger magnitude of $R_d$ seems to correspond to a larger blueshift velocity and a larger line width. By comparison, in the northern part of the main ribbon, there is much less central reversal and the Mg II k line shows mostly a redshift.

\begin{figure}
    \centering
    \epsscale{1.1}
    \plotone{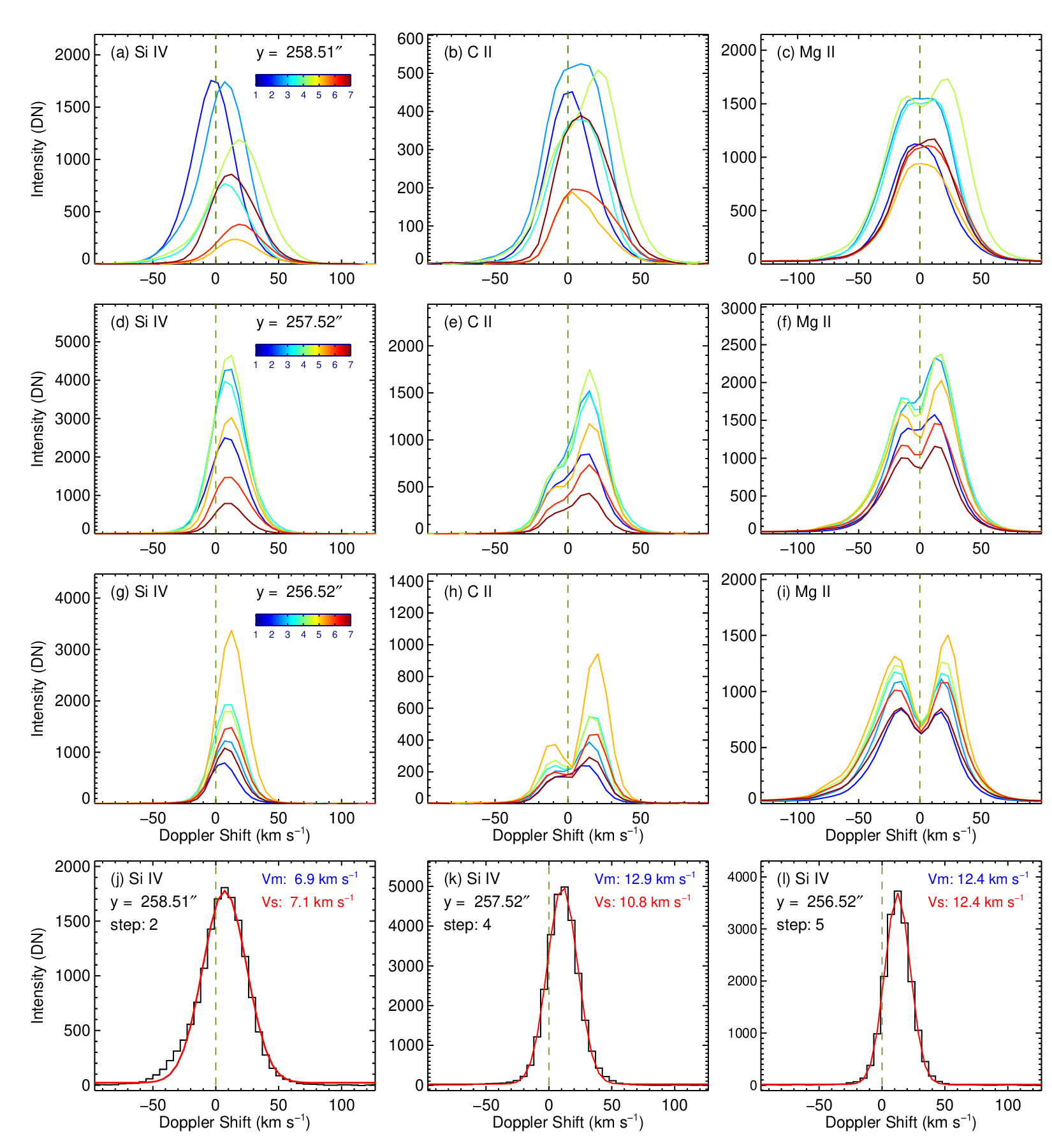}
    \caption{(a)--(i): Si IV, C II, and Mg II line profiles at the selected positions and time steps marked by the colored diamonds in Figure \ref{fig:dip_3}. Different colors represent different time steps. (j)–(l): Example profiles (black) and the single-Gaussian fittings (red) for the Si IV line at the three selected positions. The velocities by using the moment method ($V_m$) and single-Gaussian fitting ($V_s$) are given in the panels. The vertical dashed line in each of the panels denotes the reference wavelength of the spectral line.}
    \label{fig:dip_sample_3}
\end{figure}

Figure \ref{fig:dip_sample_3} shows the Si IV, C II, and Mg II line profiles at the three selected positions (marked by colored diamonds in Figure \ref{fig:dip_3}). From Figures \ref{fig:dip_sample_3}(a)--(c), we can find that at the first selected position, only the Mg II k line profile exhibits a central reversal at the fourth time step (see the green line) when the line intensity is the largest. In particular, such a central reversal of the Mg II k line is also associated with a larger redshift in the Si IV and C II lines. At this position, the Si IV and C II line profiles are both singly peaked but with a larger line width than the other two positions. From Figures \ref{fig:dip_sample_3}(d)--(f), it is obvious that at the second position, both the C II and Mg II lines show red asymmetries, while only the Mg II k line has an notable central reversal for the most time steps. As for the third position, the C II and Mg II k lines present an obvious central reversal, while the Si IV line is still singly peaked. At this position, the Mg II k line shows a more significant central reversal than the other two positions, accompanied by a blue-wing enhancement. Compared to the former two flares, the shape of the Si IV line profile in this small flare is relatively simple. As shown in Figures \ref{fig:dip_sample_3}(j)--(l), it can be fitted well with a single Gaussian function, whose Doppler velocity is much smaller than the average redshift velocities in the X1.3 and M1.5 flares.

\section{Summary and Discussions} \label{sec: sum and discuss}

In this paper, we perform a detailed analysis on the spectral features of Si IV, C II, and Mg II lines at the flare ribbons of three solar flares with different classes observed by IRIS in the sit-and-stare mode with a time resolution as high as $\sim$1 s. In particular, we quantitatively study the central reversal characteristics of the spectral line profiles at the flare ribbons. Our observational results can be summarized as follows.

\begin{enumerate}
\item [1)] At the main part of the flare ribbon, the Si IV 1402.77 \AA\ line shows different kinds of line shifts or asymmetries in dependence on the flares, like an enhanced red wing in the X1.3 flare, a red-shifted component in the M1.5 flare, and a wholly redshifted profile in the C1.6 flare. Such redshifts (asymmetries) have a lifetime of 1--3 minutes and the average velocities within the main ribbon are 56.6, 25.6, and 10.5 km s$^{-1}$ for the X-, M-, and C-class flares, respectively. It is interesting that, in the X1.3 flare, the Si IV line has an obvious central reversal at the main ribbon and several points in its front during the impulsive phase. The central reversal exists for about 10--80 s with a spatial scale of 0.5--2$^{\prime\prime}$ and an average relative depth ($R_d$) of 0.69. In addition, as the X1.3 flare evolves, the variation of $R_d$ seems to be positively correlated with the variations of line intensity and Doppler shift. However, the central reversal of the Si IV line is not detected in the M1.5 and C1.6 flares.
\item [2)] The C II 1334.53 \AA\ line shows similar line asymmetries to the Si IV line. The average redshift velocities of C II within the main ribbon are 53.2, 25.1, and 8.8 km s$^{-1}$ for the X1.3, M1.5, and C1.6 flares, respectively, which are slightly smaller than the values for the Si IV line. The central reversal of the C II line mainly appears at the leading edge and the main ribbon, which lasts for less than 2 minutes with a spatial scale of smaller than 1$^{\prime\prime}$ with a relatively dispersed distribution. The average values of $R_d$ at the main ribbon are 0.57, 0.63, and 0.48 for the X-, M-, and C-class flares, respectively. 
\item [3)] The Mg II h or k line shows more complicated line shifts (or asymmetries). Within the main part of the flare ribbons, the Mg II line shows redshifts with average Doppler velocities of 30.9, 14.4, and 2.2 km s$^{-1}$ for the X1.3, M1.5, and C1.6 flares, respectively, which are much smaller than that of the Si IV and C II lines. On the other hand, the Mg II line exhibits a blue-wing enhancement at the leading edge of the ribbons in the X1.3 and M1.5 flares and in the south part of the ribbon in the C1.6 flare, which corresponds to an upward Doppler velocity of several km s$^{-1}$. The Mg II line profiles show an obvious central reversal at the leading edge, the main ribbon, and also the trailing part in the three flares. The central reversal within the main ribbon is also accompanied by a larger relative depth ($R_d$) and the average values of $R_d$ are 0.50, 0.50, and 0.29 for the X-, M-, and C-class flares, respectively. The spatial size of the central reversal can reach up to 3.5$^{\prime\prime}$ and last for about 4 minutes.
\end{enumerate}

\begin{figure}	
    \centering
    \epsscale{1.25}	
    \plotone{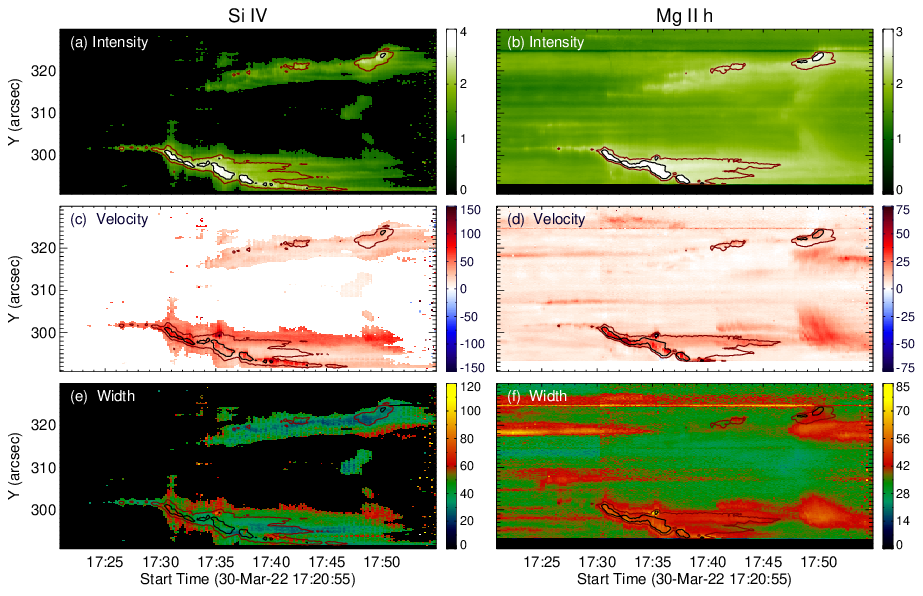} 
    \caption{Spatio-temporal variations of the total intensity (in units of log(DN)), Doppler velocity (km s$^{-1}$), and line width (km s$^{-1}$) of the Si IV and Mg II h lines along the IRIS slit obtained from the moment analysis. The contours indicate the flare regions with levels of (2$\%$, 18$\%$) and (18$\%$, 35$\%$) of the peak total intensities for the Si IV and Mg II h lines, respectively, with the black contour marking the main ribbon. Note that for the Si IV line, we set an intensity threshold of 40 DN for a reliable SNR.}
    \label{fig:momet_all_1}
\end{figure}

Theoretically, the typical duration of chromospheric condensation downflows is roughly 30 s \citep{Fisher_1989}. However, in real observations, the redshifts of chromospheric lines always persist much longer than the theoretical lifetime of chromospheric condensation, just as what is revealed in the X1.3 flare studied here. There are two possibilities to account for the long-lasting line redshifts. One is that observations have an integration effect over multiple flare loops that are heated separately. The other is that the cooling of the heated plasma from earlier loops may lead to line redshifts, as discussed by \cite{Brannon_2016_coronal_rain} and \cite{Tian_2018_evaporation}. In fact, we observed cooling plasma falling down to the ribbon from the flare loops after $\sim$17:45 UT in the X1.3 flare, as indicated by an inverse Y-shaped distribution of the line intensity, redshift, and broadening in the brightening region (Figure \ref{fig:momet_all_1}). Such redshifted line profiles caused by the cooling downflows have also been reported in \cite{2021_Panos_2021} and \cite{2020_Zhou}.

The Si IV line, which can reveal the physical properties of the TR, has been widely used in spectroscopic studies on solar explosive events, including small-scale events \citep{2014_Tian_small_scale_jets,2014Sci_Peter_hot_explosions,2017_Huang_explosive_events,2019_Chen_explosive_events} and solar flares \citep{Tian_2018_evaporation,2018_Tei,2019_Ying,2020_Yu}. In particular, the Si IV line shows different kinds of profiles in different events. In addition to single-Gaussian and double-Gaussian shapes in optically thin condition \citep{2020_Yu} during flares, \cite{2014Sci_Peter_hot_explosions} observed broadened Si IV line profiles with double-peaks combined with absorption at the line wings. Since the intensity ratio between the Si IV resonance lines is close to 2, the optically thin condition is still roughly valid and the two peaks can be well explained by a model of magnetic reconnection outflows in the lower chromosphere, rather than the self-absorption near the line center. The absorption at the line wings might be due to the absorption of the pre-existing chromosphere plasma by some singly ionized species such as Fe II and Ni II. On the other hand, \cite{2015_Yan_absorp_transient_bright} detected self-absorption in the Si IV resonance lines with the intensity ratio smaller than 2 during a transient brightening event, which was explained as the result of the overlying cool TR loops. \cite{2022_Zhou_opacity} found that most Si IV line profiles at the flare ribbons have a central reversal, which implies an opacity effect. It is interesting that although the ratio of wavelength-integrated intensities is still close to 2, the wavelength-dependent intensity ratio varies significantly from the line center to wings \citep{2022_Zhou_opacity}. \cite{2021_Panos_2021} studied a large sample of IRIS spectra and found that during the impulsive phase, the Si IV line profiles show blueshifted reversals typically at the flare ribbon front, which are ascribed to the opacity effect. Theoretically, by performing radiation hydrodynamic simulations of electron-beam-heated atmospheres, \cite{Kerr_2019} found that the Si IV resonance lines show an opacity effect especially during the heating phase, which can occur even in weaker flares for some particular beam parameters.

In our study, the Si IV 1402.77 \AA\ line profiles show an obvious central reversal only in the impulsive phase of the X1.3 flare, which are similar to the line profiles presented by \cite{2022_Zhou_opacity}. As was shown by \cite{2022_Zhou_opacity}, the intensity ratio between the Si IV resonance lines deviates from 2, which indicates that such a central reversal is mainly due to the opacity effect. Thus, it is reasonable that, in the three flares analyzed in this work, the Si IV line becomes optically thick probably in the largest flare only, especially during its impulsive phase. Interestingly, the central reversal of the Si IV line profiles occurs not only in part of the flare ribbon front, but also in the main ribbon in the X1.3 flare. During the impulsive phase of the X1.3 flare, an obvious HXR emission above 25 keV and even 50 keV was detected, implying a heating by nonthermal electrons. Considering the simulation results of \cite{Kerr_2019}, we can infer that the heating by nonthermal electrons is more significant in the X1.3 flare and plays an important role in the opacity effect of the Si IV line. In addition, the relative depth of the central reversal in the main ribbon tends to be larger, along with a lager line intensity and a red asymmetry. This is conceivable that a stronger heating could cause a more obvious opacity effect and also a stronger chromospheric condensation. However, there is no central reversal of the Si IV line in the precursor phase of the X1.3 flare or the other two flares, in which cases the heating may be not strong enough to cause noticeable opacity effect. 

The behavior of the C II resonance lines is similar to both the TR and chromospheric lines, such as the Si IV and Mg II lines, and their line width and line shift can be powerful diagnostics of the velocity field between the lower transition region and the upper chromosphere \citep{2015_Rathore_C_II_IRIS}. According to the simulations by \cite{2015_Rathore_Dig_C_II_a} and \cite{2015_Rathore_Dig_C_II_b}, the C II resonance lines are mainly formed in an optically thick condition and show double-peaked line profiles, whose source function has a local maximum in the lower chromosphere. If the source function continues to increase with height in the line forming region, the resultant line profile tends to be singly peaked. Based on the spectral observations of a solar flare, \cite{2022_Sainz_inversions_C_II_Mg_II} performed the k-means clustering with a non-local thermodynamic equilibrium (non-LTE) inversion for the Mg II and C II lines. The authors found that the centrally reversed line profiles with red asymmetries mainly appear at the leading edge of the flare ribbons \citep[see also][]{2021_Panos_2021}{}{}, which correspond to the local maximum of the temperature near the middle chromosphere. By comparison, the single-peaked profiles at the trailing region are due to the increase of the temperature in the upper chromosphere. \cite{2022_Sainz_inversions_C_II_Mg_II} also noted that the energy is initially deposited in the middle chromosphere and causes a heating there, and then the heating in the higher atmosphere becomes more pronounced as the flare evolves, which results in a transition of the spectral lines from double-peaked to single-peaked.

In the three flares here, the central reversal of the C II 1334.53 \AA\ line is distributed not only at the leading edge, but also in the main ribbon. The centrally reversed C II line profiles always have a larger peak intensity with a more obvious red asymmetry than that at the trailing part of the flare ribbons, especially in the X1.3 flare. This is similar to what is revealed by the Si IV line for the X1.3 flare. Note that the central reversal of the C II 1334.53 \AA\ line shown here is less than that for the C II 1335.71 \AA\ line \citep[see also][]{2015_Rathore_Dig_C_II_b,2015_Rathore_C_II_IRIS}{}{}, although the behavior of these two lines is always similar. The central reversal of the C II line disappears (e.g., at the trailing region) as the flare evolves, which is supposed to be due to the heating in the line formation region.

By comparison, the central reversal of the Mg II h or k line during the three flares appears in a much larger region of the flare ribbons and lasts relatively longer, indicating that the opacity effect of the Mg II line is more significant than the other two lines. The relative depth of the central reversal tends to be larger at the main ribbon, which is also related to a more obvious red asymmetry of the line profiles. Note that in the C1.6 flare, the magnitude of central reversal is relatively larger in the southern part of the ribbon where the Mg II line profiles show a blue-wing enhancement.

According to the simulation results by \cite{2013_Leenaarts_b}, the intensity of the Mg II emission peak has a positive relation with the temperature at its formation height, and the intensity ratio between the red and blue peaks correlates strongly with the average velocity in the middle chromosphere. \cite{2019_Kerr_2019_Mg_II_a} showed that an enhanced red peak of the Mg II line is due to the condensation downflows. On the other hand, \cite{2018_Panos_Mg_II} found that the broadened Mg II line profiles with a blueshifted central reversal mainly occur in the flare ribbon fronts during the impulsive phase. Such profiles are usually correlated with the nonthermal HXR emission, which indicates a heating by nonthermal electrons in the chromosphere. \cite{2023_Polito_front} quantified the spectral properties of the Mg II line at the flare ribbons and applied a simulation with nonthermal electron beam heating. They concluded that the heating at the ribbon front is more gradual with a modest energy flux, which causes a weaker chromospheric evaporation, while the heating is more impulsive and stronger in the bright ribbon. In particular, if the electron beams have a larger low-energy cutoff, the energy is mostly deposited in the deeper atmosphere, which can result in a significant central reversal of the Mg II line. Hence, we suppose that the centrally reversed Mg II line profiles with a 
large red asymmetry in the main ribbon could be the result of intense heating by nonthermal electrons with preferentially a larger low-energy cutoff.

As for the blue-wing enhancement or blue asymmetry of the Mg II line, there are several possible interpretations. One is the absorption in the red wing due to cool downflows in the upper chromospheric layer \citep{1994_Heinzel_red_absorp}, which usually causes a weaker red peak of the line profile. Another more general explanation is upward moving of the plasma in the upper chromosphere \citep{1990_Canfield_upward}. \cite{2018_Tei} observed an enhanced blue wing with, however, a weaker blue peak of the Mg II line at a flare ribbon front, which was proposed to be caused by a cool upflow overlying the chromopheric evaporation region. \cite{2019_Huang_bluewing_Mg_Ha} also observed a similar spectral feature of the Mg II line and found that it is due to an increase of the electron density and the velocity field in the line formation region. By performing radiative hydrodynamic simulations, \cite{2020_Hong_Mg_bluewing} concluded that the blue-wing enhancement of the Mg II k line originates from upward moving plasma which can be formed at different regions depending on the electron beam parameters. However, \cite{2019_Kerr_2019_Mg_II_a} suggested that the partial frequency redistribution could be important for the blue-wing enhancement of the Mg II line, which is different from the interpretations in terms of the plasma motion. 

In summary, among the three flares here, the central reversal of the Si IV line is only found in the X1.3 flare during its impulsive phase, typically in the main ribbon region. The central reversal of the C II line mainly occurs at the leading edge and the main ribbon in the three flares. By comparison, the central reversal of the Mg II line appears in a much larger region and lasts longer, indicating that the opacity effect of the Mg II line is more significant. In addition, the centrally reversed line profiles are usually associated with a larger redshift velocity in the flare main ribbon. A possible explanation is that in the flare cases analyzed here, the heating at the main ribbon is relatively stronger resulting in a stronger condensation region, which is more likely to produce the opacity effect in the Si IV line. Meanwhile, the heating from nonthermal electron beams can happen in a deeper atmosphere at the main ribbon, which causes the central reversal of the C II and Mg II lines. As the flare evolves, the atmosphere is sufficiently heated and then the optically thick lines become singly peaked. A detailed radiative hydrodynamic simulation for the three lines is required in the future to test this point.

\begin{acknowledgments} 
    We are grateful to the anonymous referee for the detailed suggestions. IRIS is a NASA Small Explorer mission developed and operated by LMSAL with mission operations executed at NASA Ames Research Center and major contributions to downlink communications funded by the Norwegian Space Center (NSC, Norway) through an ESA PRODEX contract. SDO is a mission of NASA’s Living With a Star Program. Solar Orbiter is a space mission of international collaboration between ESA and NASA, operated by ESA. The STIX instrument is an international collaboration between Switzerland, Poland, France, Czech Republic, Germany, Austria, Ireland, and Italy. The authors are supported by National Key R\&D Program of China under grants 2022YFF0503004 and 2021YFA1600504 and by NSFC under grants 12127901, 12273115, and 12233012. Y.L. is also supported by the CAS Pioneer Talents Program for Young Scientists.
\end{acknowledgments}

%\bibliography{manuscript}{}
%\bibliographystyle{aasjournal}

\end{CJK}
\end{document}